\documentstyle[11pt,aaspp4,psfig]{article}

\slugcomment{}

\def\etal{et al. }

\begin{document}

\hskip 5.truein Accepted: ApJL

\title{Abundances in Spiral Galaxies: Evidence for Primary Nitrogen Production}

\author{Liese van Zee\altaffilmark{1}}
\affil{National Radio Astronomy Observatory,\altaffilmark{2} PO Box 0, Socorro, NM 87801}
\affil{lvanzee@nrao.edu}
\author{John J. Salzer\altaffilmark{3}}
\affil{Astronomy Department, Wesleyan University, Middletown, CT 06459--0123}
\affil{slaz@parcha.astro.wesleyan.edu}
\and
\author{Martha P. Haynes}
\affil{Center for Radiophysics and Space Research}
\affil{and National Astronomy and Ionosphere Center\altaffilmark{4}}
\affil{ Cornell University, Ithaca, NY 14853}
\affil{haynes@astrosun.tn.cornell.edu}
\authoremail{lvanzee@nrao.edu}
\altaffiltext{1}{Jansky Fellow} 
\altaffiltext{2}{The National Radio Astronomy Observatory is a facility of
the National Science Foundation, operated under a cooperative agreement
by Associated Universities Inc.}
\altaffiltext{3}{NSF Presidential Faculty Fellow}
\altaffiltext{4}{The National Astronomy and Ionosphere Center is operated by Cornell
University under a cooperative agreement with the National Science Foundation.}

\begin{abstract}
We present the results of nitrogen and oxygen abundance measurements 
for 185 \ion{H}{2} regions spanning a range of radius 
in 13 spiral galaxies.  As expected, the 
nitrogen--to--oxygen ratio increases linearly with the oxygen abundance for 
high metallicity \ion{H}{2} regions, indicating that nitrogen is {\it predominately}
a secondary element.  However, the nitrogen--to--oxygen ratio plateaus for oxygen
abundances less than 1/3 solar (12+log(O/H) $<$ 8.45), as is also seen in
low metallicity dwarf galaxies.  This result suggests that the observed trend in 
dwarf galaxies is not due to outflow of enriched material in a shallow 
gravitational potential.  While the effects of
infall of pristine material and delayed nitrogen delivery are
still unconstrained, nitrogen does appear to have both a primary
and secondary component at low metallicities in all types of galaxies.
\end{abstract}


\section{Introduction}
Abundance measurements provide probes of the elemental
enrichment process and thus are important tracers of galaxy formation
and evolution.  Recently, chemodynamical models have attempted
 to obtain self--consistent star formation
histories for galaxies by combining both chemical and dynamical processes (e.g., Samland,
Hensler, \& Theis \markcite{SHT97}1997).  These models require an understanding of 
the star formation and enrichment processes, including stellar nucleosynthesis, dust depletion, 
and gas dynamics.  While stellar yields for many elements, such as O, Ne, S, and
Ar (e.g., Woosley \& Weaver \markcite{WW95}1995), are in good 
agreement with their observed enrichment ratios (e.g., Thuan, Izotov, \& Lipovetsky 
\markcite{TIL95}1995), the effective yields and origin of some elements, such
as nitrogen, are still a matter of debate.  Understanding the origin of
nitrogen is particularly important since it is frequently used to derive the 
primordial helium abundance via regression analysis of the observed enrichment of helium
and nitrogen (e.g., Pagel \& Kazlauskas \markcite{PK92}1992). In high metallicity
environments, nitrogen is believed to be synthesized during the CNO cycle in 
intermediate mass stars via a secondary process (Renzini \& Voli \markcite{RV81}1981).
On the other hand, studies of the N/O ratio in low metallicity \ion{H}{2}
regions suggest that there may also be a primary origin for nitrogen
(e.g., Edmunds \& Pagel \markcite{EP78}1978; Garnett \markcite{G90}1990; 
Thuan \etal \markcite{TIL95}1995).   In this paper, we examine the relative enrichment of 
nitrogen and oxygen in spiral galaxy \ion{H}{2} regions in order to investigate 
the origin of nitrogen in a wide range of metallicity environments.

Previous studies of chemically enriched extragalactic \ion{H}{2} regions indicate that
the effective yield of nitrogen depends on the previous enrichment of 
C and O, and thus nitrogen is considered a ``secondary'' element 
(e.g., Torres--Peimbert, Peimbert, \& Fierro
\markcite{TPPF89}1989;  Vila--Costas \& Edmunds \markcite{VE93}1993).
The signature for secondary nitrogen production is that the N/O
ratio increases linearly with O/H.  In contrast, studies of the
nitrogen enrichment in low mass dwarf galaxies indicate that the N/O
ratio is constant for oxygen abundances less than 1/3 solar
(e.g., Garnett \markcite{G90}1990; Thuan \etal \markcite{TIL95}1995;
van Zee, Haynes, \& Salzer \markcite{vHS97b}1997b).  A constant N/O ratio 
may be indicative of primary nitrogen production.  Primary nitrogen is 
defined as nitrogen produced only out of the original hydrogen and helium
in a star, either directly or through successive stages of burning.  Low
metallicity intermediate mass stars which undergo successive dredge--ups of
their enriched core are one source of primary nitrogen (Renzini \& Voli 
\markcite{RV81}1981).  Recent work on the nucleosynthesis of massive stars
suggests that primary N may also be produced in low metallicity 
massive stars via convective overshoot (Woosley \& Weaver 
\markcite{WW95}1995).  However, the signature of primary nitrogen production,
a constant N/O ratio at low metallicities, can also be caused by dynamic processes, such
as gas infall or outflow.

Standard ``closed box'' chemical enrichment models implicitly
assume that enriched materials will be retained by the system; this
assumption may not be valid in low mass systems where the gravitational 
potential well is quite shallow.  In such objects, the products of 
nucleosynthesis may be ejected from the system during supernovae explosions.
While Edmunds \markcite{E90}(1990) states that  ``simple and pure''
outflows will not affect relative enrichment ratios, such as N/O,
gas flows are likely to be neither ``simple'' nor ``pure''.  For instance,
oxygen is  predominately made in high mass stars which
undergo more violent deaths, and thus is more likely
to be removed from the galaxy.  This {\it differential} outflow
results in a decrease in the effective yield for oxygen with a 
corresponding increase in the N/O ratio.   Furthermore, additional
complications arise from possible accretion of pristine gas from
external reservoirs.  This is particularly a concern for
dwarf galaxies since many have \ion{H}{1} envelopes which
extend well beyond the optical system (e.g., van Zee, Haynes, 
\& Giovanelli \markcite{vHG95}1995).   In dwarf galaxies, the elemental 
enrichment could be diluted by accretion, resulting in lower 
effective yields.  The effect of gas infall on the N/O ratio is complex and 
highly model dependent.  For primary nitrogen, the N/O ratio will be
uneffected by dilution;  for secondary nitrogen, the inflow of pristine
material can have a wide range of effects on the N/O ratio (Edmunds \markcite{E90}1990).  

Most studies on the origin of nitrogen have concentrated on low
luminosity dwarf galaxies, where the oxygen abundance is very low.  However,
these low mass systems are particularly susceptible to both outflow
(due to their shallow gravitational potential) and inflow (from extensive
neutral gas envelopes).  We have taken advantage of the expected 
low abundance nature of outlying \ion{H}{2} regions
in spiral galaxies (e.g., Zaritsky, Kennicutt, \& Huchra \markcite{ZKH94}1994)
 to investigate whether a large contribution of
primary nitrogen is evident in \ion{H}{2} regions associated with higher mass objects.
Previous observations of \ion{H}{2} regions in spirals  (e.g.,  Vila--Costas
 \& Edmunds \markcite{VE93}1993; Thurston, Edmunds, \& Henry \markcite{TEH96}1996)
have focussed on the brighter \ion{H}{2} regions in the 
inner galaxy, which are correspondingly more enriched. 
By focussing on {\it outlying} \ion{H}{2} regions, we  
have observed spiral galaxy \ion{H}{2} regions
with metallicities as low as 1/10 solar,
comparable to typical dwarf galaxy metallicities.
By observing massive spiral galaxies, we expect the complicating
effects of gas outflow/infall to be minimized.

\section{The Data}

The full details of the observations and data reduction will be presented in
van Zee, Salzer, \& Haynes \markcite{vSH98}(1998) (hereafter, Paper II).  In
brief, optical spectroscopy of 185 \ion{H}{2} regions in 13 nearby spiral galaxies
was undertaken with the Double Spectrograph on the 5m
Palomar\footnote{Observations at the Palomar Observatory were
made as part of a continuing cooperative agreement between Cornell 
University and the California Institute of Technology.} 
telescope during 1995--1996.  During all observing runs, a
5500 \AA~dichroic was used to split the light to the two 
sides (blue and red), providing complete spectral 
coverage from 3500--7600 \AA.  The spectra were calibrated
and reddening corrected line intensities relative to H$\beta$ 
were calculated using the procedures described in van Zee,
Haynes, \& Salzer \markcite{vHS97a}(1997a).

Calculation of an elemental abundance from observed emission
line ratios requires an estimate of both the electron temperature and 
density of the \ion{H}{2} region.  Since the observed [\ion{S}{2}] 
line ratios were all within the low density limit, we
have assumed an N$_e$ of 100 cm$^{-3}$ for all regions.  The electron
temperature was either computed directly from the observed [\ion{O}{3}] 
line ratios or, if the [\ion{O}{3}] $\lambda$4363 line
was not detected, from a self--consistent temperature estimate based on an
estimate of the oxygen abundance.  An estimate of the oxygen abundance was 
obtained from the R23 calibration of McGaugh \markcite{M91}(1991); the degeneracy
between the upper and lower branches of the calibration was resolved
based on the observed [\ion{N}{2}]/H$\alpha$ line ratios.  Full details
of the oxygen abundance determination and subsequent electron temperature
estimate will be presented in Paper \markcite{vHS98}II.  Finally, the nitrogen--to--oxygen
ratio was calculated assuming that N/O = N$^+$/O$^+$ (Peimbert \& Costero \markcite{PC69}1969).
 
\section{N/O Abundance Ratios}

Figure \ref{fig:no} shows the derived N/O ratio as a function of O/H for 
the \ion{H}{2} regions in the spiral galaxy sample.  The solar value is
denoted by a filled circle (Anders \& Gervesse \markcite{AG89}1989).  
Also shown in Figure \ref{fig:no} are the results for dwarf galaxies in the 
samples of van Zee \etal \markcite{vHSB96}(1996) and 
van Zee \etal \markcite{vHS97a}(1997a)\footnote{A typographical error
was found in Table 8 of van Zee \etal (1997a) for the nitrogen abundance of UGC 5829--3; 
the correct values are: log(N/H)+12 = 6.60 $\pm$ 0.11; log(N/O)= --1.72 $\pm$ 0.12.}.  
Typical error bars for the O/H and N/O values are illustrated in
the lower right.  As expected, a linear trend is seen for the high metallicity 
\ion{H}{2} regions, indicating that secondary nitrogen production dominates in
 high metallicity environments.  At low metallicities, however, 
there is a strong deviation from the secondary nitrogen line.   Also illustrated
in Figure \ref{fig:no} is the expected trend for a combination of primary
and secondary nitrogen, assuming no time delay between the release of
oxygen and nitrogen (Vila--Costas \& Edmunds \markcite{VE93}1993).  
The area between this line and the secondary nitrogen line 
can be populated if there is a significant time delay between the delivery or
if there are dynamical effects such as outflow or infall.  With the exception of
one \ion{H}{2} region in NGC 1232, the majority of the \ion{H}{2} regions fall within 
these bounds.

It is quite clear in Figure \ref{fig:no} that the spiral galaxy and dwarf
galaxy \ion{H}{2} regions have similarly high N/O ratios at low metallicities.
Thus, in contrast to Roy \etal \markcite{RBDM96}(1996), it appears that
low mass irregulars do not have systematically lower N/O ratios than massive
disk systems.  Consequently, it is unlikely that dynamical effects, such as outflow, 
significantly affect the observed elemental abundances in dwarf galaxy samples.  
In the absence of other explanations, it seems evident that low
abundance \ion{H}{2} regions in all types of galaxies
do show evidence of primary nitrogen production.

Further support for a primary nitrogen component comes from the radial 
gradient of N/O in the spiral galaxies.  If nitrogen is purely a secondary element, 
the N/O gradient should be identical to that of O/H.  The radial gradients for
O/H and N/O are tabulated in Table \ref{tab:grad} and shown graphically in 
Figure \ref{fig:radno}.  All radii have been normalized by the isophotal
radius (R$_{25}$), as listed in the RC3 (de Vaucouleurs \etal \markcite{RC3}1991)
and tabulated in Table \ref{tab:grad}.
 In Figure \ref{fig:radno}, the filled symbols represent \ion{H}{2} 
regions from the present study.  The open circles represent data from the literature:
NGC 0628-- McCall, Rybski, \& Shields \markcite{MRS85}(1985);
NGC 1068-- Evans \& Dopita \markcite{ED87}(1987);
NGC 2403-- Garnett \etal \markcite{GSSSD97}(1997); N2903-- McCall \etal \markcite{MRS85}(1985); 
N3184-- McCall \etal \markcite{MRS85}(1985); N5457-- Kennicutt \& Garnett \markcite{KG96}(1996).  
With the exception of the literature data for NGC 1068 and NGC 2403, where the N/O ratios were 
explicitly calculated, the N/O ratios were calculated using the global relation of Thurston 
\etal \markcite{TEH96}(1996).  This relation is valid only for 8.4 $<$ 12+log(O/H) $<$ 9.2. 
It is quite clear that NGC 5457 deviates from the purely secondary trend at large
radii and low metallicity.  The solid lines in Figure \ref{fig:radno} represent the 
weighted least--squares fit for the N/O gradient. The dashed lines represent the predicted N/O 
abundances extrapolated from the O/H gradient.  In general, the N/O gradients tend to be
shallower than the O/H gradients (Table \ref{tab:grad}). For instance, the derived slope 
for N/O in NGC 2805 is significantly shallower than the O/H gradient in this galaxy.  
On the other hand, galaxies with only high metallicity \ion{H}{2} regions, such
as NGC 1637, NGC 2903 and NGC 3184, have very similar O/H and N/O gradients, as expected
for regions where secondary nitrogen production dominates.

\section{Interpretation}

We have presented clear evidence for primary nitrogen production in low 
abundance \ion{H}{2} regions in both spiral and dwarf galaxies.  At metallicities
higher than $\sim$1/3 solar, secondary nitrogen production dominates and
the signature from primary production is less evident.  Since the signature
for primary nitrogen is seen in the higher mass spiral galaxies, it is unlikely 
that dynamical processes such as infall or outflow are responsible for the constant 
N/O ratio in dwarf galaxy samples.  While there may still be some concern that even
in the spiral galaxy sample the outlying \ion{H}{2} regions could be affected by
inflow of pristine material, the evidence suggests that there is a substantial 
contribution of primary nitrogen in both dwarf and spiral galaxy \ion{H}{2} regions.

The present \ion{H}{2} region data set cannot constrain the source of the primary
nitrogen production.  However, studies of extremely low metallicity systems can
exploit the time delay between the release of oxygen and nitrogen to investigate 
in which types of stars primary nitrogen is produced.  If nitrogen is primarily
made in high mass stars (mechanism described in  Woosley \& Weaver \markcite{WW95}1995), 
there will be no time delay between the release of N and O, so the
N/O ratio will have a small scatter at low abundances.  On the other hand, if
nitrogen is predominately made in intermediate mass stars (Renzini \& Voli \markcite{RV81}1981),
there will be a significant time delay between the release of N and O.  In this
case, the N/O ratio will change as a function of time, increasing the observed
scatter (e.g., Garnett \markcite{G90}1990).  Searches for such a signature
in \ion{H}{2} region abundances have been inconclusive.  For instance,
while Thuan \etal \markcite{TIL95}(1995) find an extremely small scatter in the N/O ratio
of blue compact dwarf galaxies, other studies of \ion{H}{2} regions in
dwarf irregular galaxies result in a statistically significant scatter 
(e.g., Garnett \markcite{G90}1990; Skillman, Bomans, \& Kobulnicky \markcite{SBK97}1997;
van Zee \etal \markcite{vHS97a}1997a).
Studies of extremely low metallicity systems are necessary to resolve
this issue since the time delay between oxygen and nitrogen enrichment
will have the largest effect in such objects.

Low abundance systems are extremely rare in the local universe;  
currently the most extreme objects are I~Zw~18 (Skillman \& Kennicutt \markcite{SK93}1993)
and SBS 0335-052 (Izotov \etal \markcite{ILCFGK97}1997), 
at 1/50th and 1/40th of solar, respectively. 
 At high redshift, however, low abundance systems
are quite common. Thus, investigation of the nitrogen abundance in
high redshift damped Ly$\alpha$ systems can provide
additional constraints on the origin of nitrogen.  While this is still a
relatively new endeavor, preliminary results suggest that the N/O ratio
is {\it lower} in low metallicity damped Ly$\alpha$ systems than
in comparable low metallicity \ion{H}{2} regions (e.g., Pettini, Lipman, \& Hunstead
\markcite{PLH95}1995; Lu, Sargent, \& Barlow \markcite{LSB98}1998).
Further, the scatter increases at low metallicities in these studies.
Thus, the high redshift observations appear to be catching systems during the
time delay between the release of oxygen and nitrogen.  If this is the
case,  it is likely that primary nitrogen is predominantly produced in 
low metallicity intermediate mass stars.  Further studies of low metallicity
systems at high and low redshift will be needed to confirm these results.

\acknowledgments

We acknowledge the financial support by NSF grants AST95--53020 to JJS and AST90--23450 and AST95-28860 to MPH.

\tablenum{1}
\begin{deluxetable}{rccccccc}
\tablecaption{Radial Abundance Gradients \label{tab:grad}}
\tablehead{
\colhead{}& \colhead{Morph.\tablenotemark{a}}& \colhead{$R_{25}$\tablenotemark{a}}& \colhead{(O/H)}& \colhead{(N/O)} &\multicolumn{3}{c}{Number of HII Regions} \\
\colhead{Object}& \colhead{Type}&\colhead{[arcsec]} & \colhead{[dex/$R_{25}$]} & \colhead{[dex/$R_{25}$]} & \colhead{(O/H)} & \colhead{(N/O)}& \colhead{(New)} }
\startdata
NGC 0628  & Sc  & 314. & --0.99 $\pm$ 0.14 & --0.57 $\pm$ 0.12 & 26 & 25 & 19 \nl
NGC 0925  & Sd  & 314. & --0.45 $\pm$ 0.08 & --0.34 $\pm$ 0.04 & 53 & 44 & 44 \nl
NGC 1068  & Sb  & 212. & --0.30 $\pm$ 0.07 & --0.01 $\pm$ 0.09 & 13 &  6 &  1 \nl
NGC 1232  & Sc  & 222. & --1.31 $\pm$ 0.20 & --0.32 $\pm$ 0.10 & 16 & 16 & 16 \nl
NGC 1637  & Sc  & 120. & --0.37 $\pm$ 0.14 & --0.30 $\pm$ 0.17 & 15 & 15 & 15 \nl
NGC 2403  & Scd & 656. & --0.77 $\pm$ 0.14 & --0.40 $\pm$ 0.08 & 40 & 26 & 17 \nl
NGC 2805  & Sd  & 189. & --1.05 $\pm$ 0.17 & --0.29 $\pm$ 0.06 & 17 & 17 & 17 \nl
IC 2458   & I0  & \nodata & \nodata & \nodata &  \nodata & \nodata &  3 \nl
NGC 2820  & Sb  & \nodata & \nodata & \nodata &  \nodata & \nodata &  4 \nl
NGC 2903  & Sbc & 378. & --0.56 $\pm$ 0.09 & --0.57 $\pm$ 0.17 & 36 & 13 &  9 \nl
NGC 3184  & Scd & 222. & --0.78 $\pm$ 0.07 & --0.77 $\pm$ 0.12 & 32 & 22 & 17 \nl
NGC 4395  & Sm  & 395. & --0.32 $\pm$ 0.19 & --0.04 $\pm$ 0.11 & 14 & 10 & 10 \nl
NGC 5457  & Scd & 865. & --1.52 $\pm$ 0.09 & --0.65 $\pm$ 0.09 & 53 & 45 & 13 \nl
\enddata
\tablenotetext{a}{Morphological type and isophotal radius from RC3}
\end{deluxetable}

\vfill
\eject

\psfig{figure=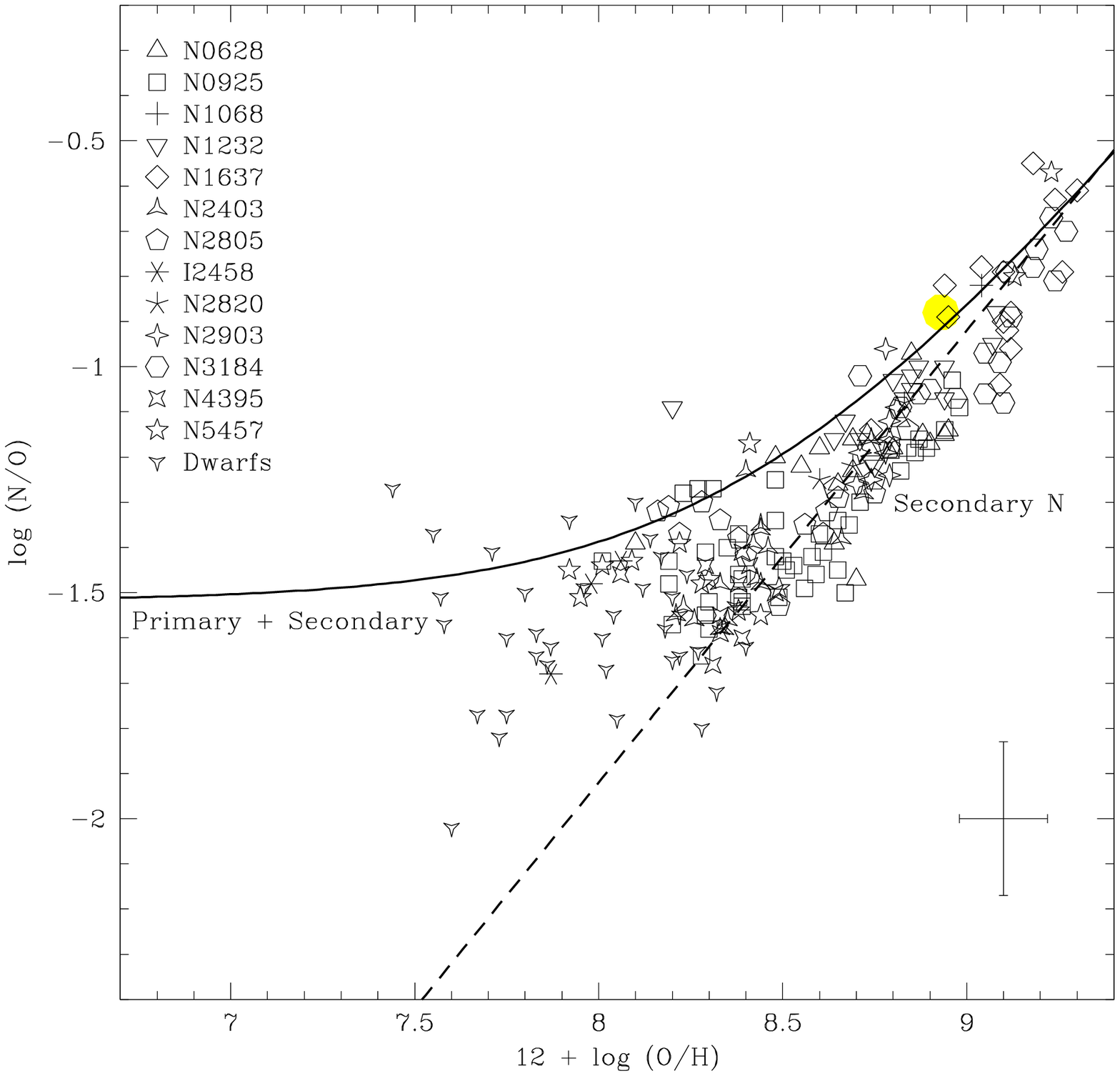,width=7.in,bbllx=1pt,bblly=150pt,bburx=600pt,bbury=700pt,clip=t}
\vskip -0.2 truein
\figcaption[N/O Plot]
{ The derived N/O ratio as a function of oxygen abundance
for \ion{H}{2} regions in spiral galaxies and dwarf galaxies (van Zee \etal 1997a);
the solar value is denoted by a filled circle (Anders \& Gervesse 1989). 
The outermost, lowest metallicity  spiral galaxy \ion{H}{2}
regions have similar N/O ratios to those found in the dwarf galaxies.  A distinct ``knee''
is seen at 12+log(O/H) of 8.45 ($\sim$1/3 of solar).  A theoretical curve for
primary and secondary nitrogen production (Vila--Costas \& Edmunds 1993) is shown.
\label{fig:no} }

\psfig{figure=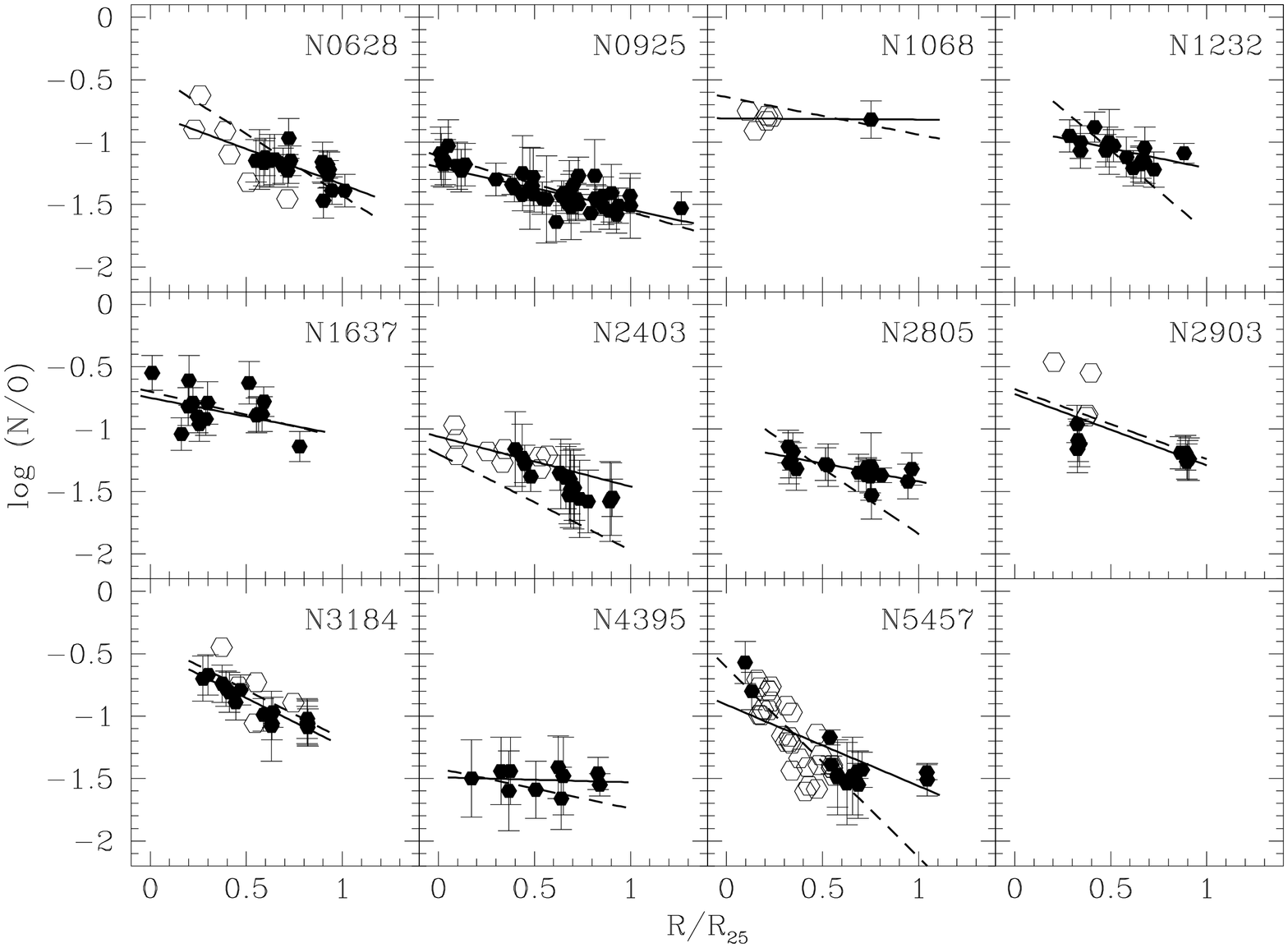,width=7.in,angle=-90.}
\figcaption[N/O Radial]
{ The N/O ratio vs. the normalized isophotal radius.
The filled symbols represent \ion{H}{2} regions from the present study. 
The open circles represent data from the literature:
NGC 0628-- McCall \etal (1985); NGC 1068-- Evans \& Dopita (1987);
NGC 2403-- Garnett \etal (1997); N2903-- McCall \etal (1985); 
N3184-- McCall \etal (1985); N5457-- Kennicutt \& Garnett (1996). The solid lines
are the derived N/O gradients; the dashed lines represent the predicted N/O
gradients if only secondary nitrogen is produced.  
\label{fig:radno} }

\end{document}